\documentclass[preprint]{elsarticle}

\usepackage{float}
\usepackage[pagewise]{lineno}
\usepackage[dvipsnames]{xcolor}
\usepackage{framed}
\usepackage{multicol}
\usepackage{amssymb}
\usepackage{todonotes}     
\usepackage[utf8]{inputenc}
\usepackage[ruled, vlined]{algorithm2e}
\usepackage{graphicx}
\usepackage{subcaption}
\usepackage{tabularx}
\usepackage{amsmath}
\usepackage{multirow}
\usepackage{setspace}
\usepackage{tabu}
\usepackage{tcolorbox}

\usepackage[percent]{overpic}

\usepackage{mylines}
\usepackage{myplots}
\newcommand{\lsy}[3]{\mbox{(\kern-.4em\lineSymbolRGB[#3]{#1}{.8pt}{#2}{4pt}\kern-.4em)}}
\newcommand{\sy}[2]{\mbox{(\kern-.25em\SymbolRGB[solid]{#1}{.8pt}{#2}{4pt}\kern-.25em)}}
\newcommand{\lcap}[2]{~\,{\kern-1em\protect\mylcap{#1}{#2}}}

\definecolor{conv}{HTML}{ffffbf}
\definecolor{pool}{HTML}{789ca5}
\definecolor{fc}{HTML}{b7ebeb}
\definecolor{tanh}{HTML}{a5a54c}
\definecolor{upsampl}{HTML}{ef9351}
\definecolor{reshape}{HTML}{ad44b6}

\DeclareMathAlphabet{\mathcal}{OMS}{cmsy}{m}{n}
\DeclareMathAlphabet\mathbfcal{OMS}{cmsy}{b}{n}

\usepackage{siunitx}
\usepackage[colorlinks=true]{hyperref}
\usepackage[noabbrev]{cleveref}

\usepackage[normalem]{ulem}
\usepackage{color}

\graphicspath{{Figs/}}

\journal{}
\begin{document}
\hypersetup{
    urlcolor=black,
    citecolor=blue,
    linkcolor=black
    }
\begin{frontmatter}
\title{Towards extraction of orthogonal and parsimonious non-linear modes from turbulent flows}

\author[1]{Hamidreza Eivazi\corref{cor}}
\ead{hamidre@kth.se}

\author[2]{Soledad Le Clainche}

\author[3]{Sergio Hoyas}

\author[1,4]{Ricardo Vinuesa\corref{cor}}
\ead{rvinuesa@mech.kth.se}

\address[1]{FLOW, Engineering Mechanics, KTH Royal Institute of Technology, \\ SE-100 44 Stockholm, Sweden}
\address[2]{School of Aerospace Engineering, Universidad Polit\'ecnica de Madrid, 28040 Madrid, Spain}
\address[3]{Instituto Universitario de Matem\'atica Pura y Aplicada, Universitat Polit\`ecnica de Val\`encia, 46022 Valencia, Spain}
\address[4]{Swedish e-Science Research Centre (SeRC), Stockholm, Sweden}


\cortext[cor]{Corresponding author}

\begin{abstract}
We propose a deep probabilistic-neural-network architecture for learning a minimal and near-orthogonal set of non-linear modes from high-fidelity turbulent-flow-field data useful for flow analysis, reduced-order modeling, and flow control. Our approach is based on $\beta$-variational autoencoders ($\beta$-VAEs) and convolutional neural networks (CNNs), which allow us to extract non-linear modes from multi-scale turbulent flows while encouraging the learning of independent latent variables and penalizing the size of the latent vector. Moreover, we introduce an algorithm for ordering VAE-based modes with respect to their contribution to the reconstruction. We apply this method for non-linear mode decomposition of the turbulent flow through a simplified urban environment, where the flow-field data is obtained based on well-resolved large-eddy simulations (LESs). We demonstrate that by constraining the shape of the latent space, it is possible to motivate the orthogonality and extract a set of parsimonious modes sufficient for high-quality reconstruction. Our results show the excellent performance of the method in the reconstruction against linear-theory-based decompositions. Moreover, we compare our method with available AE-based models. We show the ability of our approach in the extraction of near-orthogonal modes that may lead to interpretability.
\end{abstract}


\end{frontmatter}

\section{Introduction}
\label{sec:introduction}

Analysis of turbulent flows is challenging due to the presence of a vast range of spatio-temporal coherent structures in high-dimensional non-linear dynamics. However, the fact that common flow features appear across a wide range of fluid flows suggests that there are key dominant phenomena that serve as the foundation of many flows. Modal-decomposition techniques offer methods to identify a low-dimensional coordinate system for capturing dominant flow features \citep{Taira_2017,Taira_2020} useful for developing reduced-order models, analyzing non-linear and chaotic dynamics, and designing efficient flow-control schemes. Proper-orthogonal decomposition (POD) \citep{Lumley1967} and dynamic-mode decomposition (DMD) \citep{Rowley_2009,Schmid2010} are two mode-decomposition methods based on linear algebra that have been widely used to extract the dominant spatio-temporal features in fluid flows. Balanced POD (BPOD) \citep{BPOD}, spectral POD (SPOD) \citep{SPOD}, higher-order DMD (HODMD) \citep{HODMD} and spatio-temporal Koopman decomposition (STKD) \citep{STKD} are several successful variants of POD and DMD for analysis of turbulent flows.

Recent developments in deep learning for engineering problems bring advanced and innovative approaches to improve the efficiency, flexibility, and accuracy of the predictive models. Some of the outstanding applications of deep neural networks (DNNs) in the domain of computational physics are solution of partial differential equations (PDEs) \citep{Raissi2019}, operator learning \citep{gin2020,Lu2021,Zongyi2021}, linear embedding of non-linear dynamics \citep{Lusch2018}, and model reduction of dynamical systems \citep{Lee2020}. Fluid mechanics has been one of the active research topics for development of innovative DNN-based approaches \citep{kutz_2017,duraisamy_et_al,brunton_et_al_2020}. Successful application of DNNs has been shown in the data-driven turbulence closure modeling \citep{ling2016,ml_rans}, prediction of temporal dynamics of a low-order model of turbulence \citep{srinivasan,EIVAZI_2021}, extraction of turbulence theory for two-dimensional decaying isotropic turbulent \citep{jimenez_2018}, non-intrusive sensing in turbulent flows \citep{guastoni_2020,guemes2021}, and active flow control through deep reinforcement learning \citep{RL_2020}.

Besides the aforementioned linear methods for modal decomposition of flow-field data, DNN-based models have shown promising performance in learning a compact latent representation of high-dimensional data by accounting for the non-linearity in the low-dimensional mapping using non-linear activation functions \citep{Hinton_2006}. In particular, unsupervised learning based on autoencoders (AEs) has been shown suitable for efficient mode decomposition and reduced-order modeling with superior performance in flow reconstruction over the linear POD \citep{MILANO_2002,Eivazi_2020}. 

Moreover, convolutional neural networks (CNNs) \citep{Lecun_1998} and their ability in pattern recognition have received increasing attention by the fluid-mechanics community \citep{lee_you_2019,fukami_fukagata_taira_2019,kim_lee_2020,kim_kim_won_lee_2021,guastoni_2020}.~\cite{murata_fukami_fukagata_2020} proposed a CNN-based autoencoder architecture for decomposition of flow-fields into non-linear low-dimensional modes and to visualize each mode. They applied this so-called mode-decomposing convolutional-neural-network autoencoder (MD-CNN-AE) to a relatively simple laminar flow around a circular cylinder at $Re_D = 100$ (where $Re_D$ is the Reynolds number based on freestream velocity and cylinder diameter). Their results showed the superior performance of the CNN-based autoencoder over POD where the reconstruction of the flow from only two MD-CNN-AE modes contains also the higher-order POD modes. The architecture of AE-based methods allows a non-linear low-dimensional mapping leading to a superior performance against linear-theory-based methods. However, the AE-based methods do not benefit from the useful properties of the eigenvalue or singular-value-decomposition techniques, e.g., optimality and orthogonality. In contrast to the POD modes, which are an orthogonal set of basis vectors arranged in the order of their energy content, the AE-based modes are neither orthogonal nor ranked. This may lead to the lack of interpretability and robustness of the AE-based modes. In order to obtain ranked modes,~\cite{Fukami_2020} proposed a hierarchical CNN-AE architecture inspired by the concept of hierarchical autoencoder (AE) \citep{SAEGUSA_2004}. The proposed method was first applied to a laminar cylinder wake and its transient process and further to an in-plane cross-sectional velocity field of turbulent channel flow at $Re_{\tau} = 180$ (note that $Re_{\tau}$ is the friction Reynolds number, based on channel half height and friction velocity). They showed that the hierarchical autoencoder (AE) can rank the AE modes following their contributions to the reconstructed field while achieving efficient order reduction. However, issues related to interpretability and non-uniqueness remained unanswered.

In this paper, we propose a probabilistic method based on $\beta$-variational autoencoders ($\beta$-VAEs) \citep{Higgins_2017} and CNNs in order to extract a minimal (parsimonious) set of near-orthogonal non-linear modes from turbulent flows. We applied the proposed machine-learning method for the modal decomposition of high-fidelity turbulent flow simulation data of a simplified urban environment. The flow simulation is carried through well-resolved large-eddy simulation (LES) by means of the spectral-element code Nek5000 \citep{nek5000-web-page}. The results from the proposed method are compared with the results from a conventional CNN-AE model, a hierarchical CNN-AE model, and the POD. Through the training process of the available AE-based modal-decomposition methods, the objective is to only minimize the reconstruction loss. In contrast, we also minimize the correlation between the latent variables and penalize the size of the latent vector. By solving this multi-objective optimization problem using our CNN--$\beta$VAE approach, we seek a minimal set of uncorrelated non-linear modes that are able to accurately describe the turbulent flow-field data. The obtained modes are extremely useful for the development of reduced-order surrogate models or designing flow-control strategies. Moreover, understanding the physics of the turbulent flow through the extraction of uncorrelated non-linear mechanisms is another incentive for applying CNN-$\beta$VAEs for modal decomposition. In particular, the development of accurate predictive models and understanding flow structures in urban environments are of significant importance due to their impact on urban planning, air quality management, and pollutant dispersion~\citep{Vinuesa_2015}. The essential need to address sustainable development from an urban perspective is pronounced in the 2030 Agenda \citep{UNGA}\footnote{United Nations (UN)} through the sustainable development goals (SDGs) 11 (on sustainable cities and communities) and 13 (on climate action). Air pollution is a major cause of premature death and disease and is the single largest environmental health risk in Europe. Heart disease and stroke are the most common reasons for premature deaths attributable to air pollution, followed by lung diseases and lung cancer \citep{air_eu_2020}. Predictive models for air-quality detection are in particular important to provide protection from excessive pollution concentrations. However, the available predictive models are unable to provide the spatial and temporal accuracy required for reproducing pollutant-dispersion patterns within urban environments. Therefore, there is a pressing need for improved prediction and assessment methods to tackle these challenges and enable urban sustainability in the near future~\citep{VinuesaNat}. Furthermore, it is necessary to gain further insight into the physics responsible for the pollutant-concentration and thermal distributions within cities~\citep{Torres2021}.

This article is organized as follows: in $\S$\ref{sec:modal-decomposition} we provide an overview of the flow physics and discuss the theoretical background relevant to linear-theory-based modal decomposition; in $\S$\ref{sec:CNN-based-autoencoders} we discuss three different CNN-AE-based methods from a methodological point of view; the performance of the methods and their characteristics are compared in $\S$\ref{sec:results-and-discussion}; and finally, in $\S$\ref{sec:conclusions} we provide a summary and the conclusions of the study.

\section{Modal decomposition}
\label{sec:modal-decomposition}

Modal decomposition is a mathematical method to extract energetically- and dynamically-important features from fluid flows. The spatial features are represented by a set of modes ranked in terms of kinetic energy or their largest amplitude (connected with the norm of the mode). These modes are generally obtained via solving an eigenvalue problem. The obtained eigenvalues could represent the energy content of the mode or the growth rates and frequencies modelling the temporal dynamics of the flow motion. Modal-decomposition techniques produce a low-dimensional coordinate system for capturing dominant flow structures. These dominant flow structures are extremely useful not only for flow analysis but also for reduced-order modeling and flow control. For a detailed discussion of the modal-decomposition techniques, the readers are referred to the reviews on the topic \citep{Taira_2017,Taira_2020}. In the following, we discuss flow physics in a simplified urban environment and apply POD for modal decomposition.

\subsection{Training data}

We employ a database~\citep{Torres2021,TorresMS} of the flow through a simplified urban environment to obtain the training data. This database was obtained through well-resolved large-eddy simulation (LES) using the open-source numerical code Nek5000 \citep{nek5000-web-page}, which is based on the spectral-element method (SEM), to solve the incompressible Navier--Stokes equations:
\begin{equation}
    \begin{aligned}
        & \nabla \cdot \mathbf{u} = 0,\\
        & \dfrac{\partial \mathbf{u}}{\partial t} + (\mathbf{u} \cdot \nabla)\mathbf{u} = - \nabla p + \nu \nabla^2 \mathbf{u},
    \end{aligned}
\end{equation}
where $\mathbf{u}$ represents the velocity vector,  $t$ is the time, and $\nu$ is the kinematic viscosity. $p$ is the pressure, including the constant density. Note that the SEM combines the geometrical flexibility required to discretize the urban geometry with the high-order accuracy of spectral methods. The geometry of the buildings is discretized using hexahedral elements and the scale disparity of the turbulent flows is resolved using the Gauss--Lobatto--Legendre (GLL) quadrature within each element. The flow case includes two wall-mounted obstacles with width-to-height ratio $b/h$ and height-to-separation ratio $h/\ell$ of 0.5 and 1.25, respectively. Here $x$, $y$ and $z$ denote streamwise, wall-normal and spanwise directions, respectively. The $x$--$z$ cross-sectional velocity fields at $y = 0.5h$ are extracted and used as the input data. As we focus on the flow around the obstacles, we extract the following region of the computational domain: $-1h \leqslant x \leqslant 5h$ and $-1.5h \leqslant z \leqslant 1.5h$. To facilitate the data-handling process, we performed spectral interpolation of the 1,200 instantaneous fields considered in this work from the original SEM mesh to another one containing the following number of grid points in $x$ and $z$: $(N_x, N_y) = (96, 192)$. We use the fluctuation component of the streamwise velocity $u$ as the input of the models. 
  
\subsection{Proper-orthogonal decomposition (POD)}
\label{sec:POD}

In this section, we apply POD for modal decomposition and discuss some of its useful properties, i.e., optimality and orthogonality. The POD technique (also known as the Karhunen-Lo\`eve (KL) procedure \citep{karhunen1946,loeve1955}) was first introduced to the fluid dynamics/turbulence community by \cite{Lumley1967} as a mathematical algorithm to extract coherent structures from turbulent flows. POD extract modes based on minimizing the mean-square error between the signal and its reconstruction and also minimizing the number of modes required for such a reconstruction. This leads to a minimal number of basis functions or modes to capture as much energy as possible.
Let us consider a vector field as: $\boldsymbol{q}(\boldsymbol{\xi}, t)$ representing, e.g., velocity with coordinates $\boldsymbol{\xi}$, where $t$ denotes time. Having the temporal mean $\overline{\boldsymbol{q}}(\boldsymbol{\xi})$ subtracted, the unsteady component of the vector field can be decomposed as: 
\begin{equation}
    \boldsymbol{q}(\boldsymbol{\xi}, t) - \overline{\boldsymbol{q}}(\boldsymbol{\xi}) = \sum_j a_j(t)\boldsymbol{\phi}(\boldsymbol{\xi}),
\end{equation}
where $\boldsymbol{\phi}(\boldsymbol{\xi})$ and $a_j$ represent the spatial modes and temporal (expansion) coefficients respectively. To this end, we first prepare snapshots of the flow-field as a collection of finite-dimensional data vectors:
\begin{equation}
    \boldsymbol{x}(t) = \boldsymbol{q}(\boldsymbol{\xi}, t) - \overline{\boldsymbol{q}}(\boldsymbol{\xi}) \in \mathbb{R}^n,~~~~t = t_1, t_2, \dots, t_m.
\end{equation}
where $\boldsymbol{x}(t)$ represents the fluctuating component of the vector data, $n$ is the number of grid points modelling the vector data and $m$ is the number of snapshots selected to model the flow dynamics. We arrange the data into a matrix $\boldsymbol{X}$ through the concatenation of $m$ snapshots as follows:
\begin{equation}
    \boldsymbol{X} = [\boldsymbol{x}(t_1)~\boldsymbol{x}(t_2)~\dots~\boldsymbol{x}(t_m)]\in \mathbb{R}^{n\times m}.
\end{equation}
The POD modes can be determined as the eigenvectors of the covariance matrix $\boldsymbol{R} = \boldsymbol{XX}^\mathsf{T}$:
\begin{equation}
    \boldsymbol{R}\boldsymbol{\phi}_j = \lambda_j \boldsymbol{\phi}_j,~~~~\boldsymbol{\phi}_j \in \mathbb{R}^n,~~~~\lambda_1\geq\cdots\lambda_n\geq 0.
\end{equation}
The eigenvalues $\lambda_j$ show how well each mode $\boldsymbol{\phi}_j$ represents the reference data in the $\ell_2$ sense. Considering the velocity vectors as $\boldsymbol{x}(t)$, each eigenvalue shows the kinetic energy captured by its corresponding mode. 

Another approach is to apply singular-value decomposition (SVD) \citep{Sirovich_1987} directly on the matrix $\boldsymbol{X}$ as:
\begin{equation}
    \boldsymbol{X} = \mathbf{\Phi \Sigma \Psi}^\mathsf{T},
\end{equation}
where $\mathbf{\Phi} \in \mathbb{R}^{n \times n}$ and $\mathbf{\Psi} \in \mathbb{R}^{m \times m}$ are the left and right singular vectors of $\boldsymbol{X}$, respectively, and $\mathbf{\Sigma}$ is a diagonal matrix containing the singular values $(\sigma_1, \dots, \sigma_m)$. The singular vectors $\mathbf{\Phi}$ and $\mathbf{\Psi}$ are identical to the eigenvectors of $\boldsymbol{XX}^\mathsf{T}$ and $\boldsymbol{X}^\mathsf{T}\boldsymbol{X}$, respectively, and the singular values are related to the eigenvalues by $\sigma_j^2 = \lambda_j$. Two important properties of the POD are:
\begin{itemize}
    \item Optimality: This indicates that the POD is the most efficient decomposition, in the sense that for a given number of modes $k$, the projection on the subspace spanned by the leading $k$ modes contains the greatest possible energy on average among all linear decompositions.
    \item Orthogonality: This implies that the time series of the coefficients $a_j(t)$ are linearly uncorrelated, which is an attractive property for constructing reduced-order models. 
\end{itemize}
It is important to note that the POD is a linear procedure. Linearity is the origin of the strength of the method, its applicability, but it is also its limitation. As stated above regarding the optimality of the POD results, it should be noted that optimality is implied only with respect to other linear representations \citep{holmes_1996}.

\begin{figure}[h]
    \centering
    \includegraphics[width=\textwidth]{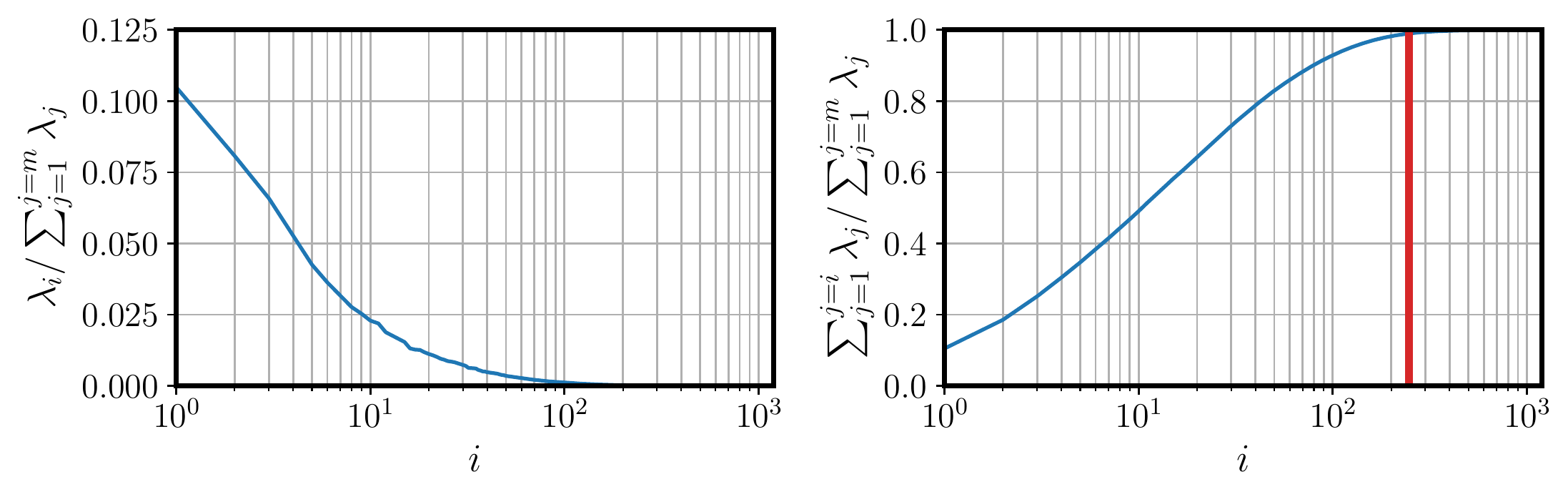}
    \caption{Eigenvalues $\lambda_i$ (left) and the cumulative eigenvalue spectrum $\sum_{j = 1}^{j = i} \lambda_j$ (right) normalized with the cumulative sum of the eigenvalues $\sum_{j = 1}^{j = m} \lambda_j$, where $i$ indicates the number of modes. The solid red line shows the number of modes required to capture 99\% of the energy.}
    \label{fig:eigenvalues}
\end{figure}

We applied POD using the SVD method on the urban-flow database discussed above. \Cref{fig:eigenvalues} shows the eigenvalues $\lambda_i$ (left) and the cumulative eigenvalue spectrum $\sum_{j = 1}^{j = i} \lambda_j$ (right) normalized with the cumulative sum of the eigenvalues $\sum_{j = 1}^{j = m} \lambda_j$, where $i$ indicates the number of modes. We observed that 247 modes are required to capture 99\% of the energy as it is depicted by the vertical red line in \cref{fig:eigenvalues} (right). This result implies that it is impractical to represent turbulent flows as a linear superposition of a few modal functions, and thus, more sophisticated algorithms enabling a non-linear modal decomposition are required.

\section{CNN-based autoencoders for modal decomposition}
\label{sec:CNN-based-autoencoders}

An autoencoder is a deep neural network (DNN) with an architecture suitable for unsupervised feature extraction. The network comprises of two parts: an encoder that maps the input data to a low-dimensional latent space $\boldsymbol{x} \mapsto \boldsymbol{r}$, and a decoder that projects the latent vector $\boldsymbol{r}$ back to the reference space $\boldsymbol{r} \mapsto \boldsymbol{\tilde{x}}$. We refer to the encoder and decoder parts as $\mathcal{E}$ and $\mathcal{D}$, respectively. Through the model training, the autoencoder learns to extract the most important features in the data that are required for reconstruction by optimizing the model parameters $\boldsymbol{w}$ to minimize the reconstruction loss $\mathcal{L}_\mathrm{rec}$:
\begin{subequations}
    \begin{equation}
        \mathcal{F} = \mathcal{D} \circ \mathcal{E}, 
    \end{equation}
    \begin{equation}
        \boldsymbol{\tilde{x}} = \mathcal{F}(\boldsymbol{x}; \boldsymbol{w}), \\
    \end{equation}
    \begin{equation}
        \mathcal{L}_\mathrm{rec} = \epsilon(\boldsymbol{x}, \boldsymbol{\tilde{x}}), 
    \end{equation}
\end{subequations}
where $\boldsymbol{x}$ denotes the input data, which in our case is the fluctuating component of the streamwise flow velocity; $\boldsymbol{\tilde{x}}$ is the reconstruction of the input data and $\epsilon$ represents the loss function. The autoencoder architecture is attractive for modal decomposition as it provides a framework that can incorporate non-linearity in the mappings through the use of non-linear activation functions. 

Another challenge in the modal decomposition of turbulent flows is the process of information from input fields that contain multiscale coherent features. The presence of coherent features motivates the use of convolutional layers in autoencoder models to process the input information. Through a two-dimensional convolution layer and using a so-called filter with size $H\times H \times K$, the data is processed as: 
\begin{equation}
    \boldsymbol{z}^{(l)}_{ijm} = \varphi \left ( \sum_{k = 1}^{K} \sum_{p = 1}^{H} \sum_{q = 1}^{H} \boldsymbol{z}^{(l-1)}_{i+p, j+q, k}\boldsymbol{w}^{(l)}_{pqkm} + \boldsymbol{b}^{(l)}_{ijm} \right ),
\end{equation}
where $\boldsymbol{z}^{(l-1)}$ and $\boldsymbol{z}^{(l)}$ represent the input and output variables, respectively. Note that $\boldsymbol{w}^{(l)}$ and $\boldsymbol{b}^{(l)}$ denote the weights and the biases of layer $l$, and $\varphi$ denotes the activation function. Furthrmotr, $K$ is the number of the so-called kernels, which are the two-dimensional slices of the filter. Since $H \ll N_x,N_y$, the use of kernels greatly reduces the number of parameters that need to be learned during training. It is common to use convolution layers together with the pooling and upsampling layers. Through a pooling operation, the data are compressed by a factor of $(1/P)^2$ so that a region with size of $P \times P$ is represented by e.g. its maximum value; this is the so-called max-pooling operation. The upsampling operation is used to expand the data dimension by e.g. nearest-neighbor or bilinear interpolation. In the following, we discuss two NN-architectures based on CNNs and autoencoders, namely CNN-based autoencoders (CNN-AEs) and CNN-based hierarchical autoencoders (CNN-HAEs). Finally, we propose CNN-based $\beta$-variational autoencoders (CNN-$\beta$VAEs) for non-linear modal decomposition of turbulent flows. For all the models, we use mean-squared error as the loss function for reconstruction $\mathcal{L}_{\mathrm{rec}}$ and the Adam algorithm \citep{kingma2017adam} to optimize the model parameters $\boldsymbol{w}$. We employ the early-stopping criterion and obtain the best model based on the validation loss to avoid overfitting, where 20\% of the data is randomly selected for validation.

\subsection{CNN-based autoencoders}

\begin{figure}[h]
    \centering
    \includegraphics[width=\textwidth]{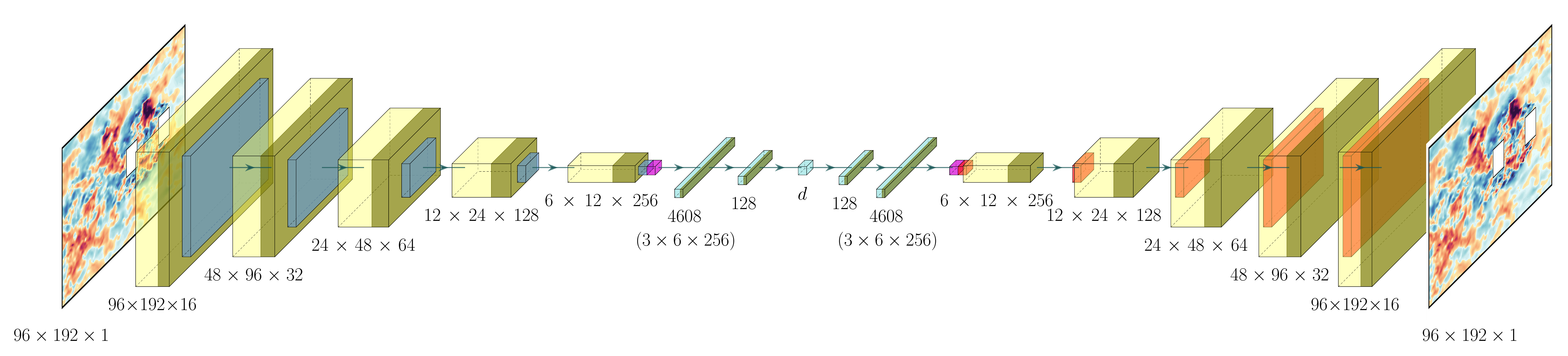}
    \caption{Schematic view of the CNN-AE architecture. The color coding for each layer is: 2D-convolution \sy{conv}{b}, tanh activation \sy{tanh}{b}, max pooling \sy{pool}{b}, reshape \sy{reshape}{b}, fully-connected layer \sy{fc}{b}, upsampling \sy{upsampl}{b}.}
    \label{fig:sch_AE}
\end{figure}

\Cref{fig:sch_AE} depicts a schematic representation of the CNN-AE model. For simplicity, consider the fluctuating component of the streamwise velocity $u$ as the input/output of the model, but it is also possible to consider all the velocity components ($u$, $v$, $w$) as the input/output. The first convolution layer contains 16 filters with a size of $(3 \times 3)$, and it is followed by a max-pooling layer with $P = 2$. At each convolution step, we double the number of feature maps to extract more information from the turbulent-flow data while at each downsampling step we reduce the dimension. This allows the next layer to combine the features individually identified in each feature map, enabling the extraction of larger and more complex features for progressively deeper convolutional networks from simple non-linear combinations of the previous ones. Therefore, convolutional layers can learn to recognize turbulent-flow patterns of various complexity and scales~\citep{guastoni_2020}. After five steps of convolution and max pooling, the extracted features are flattened and fed to fully connected layers to reduce the dimension to the latent vector $\boldsymbol{r}$ with a size of $d$. The latent vector $\boldsymbol{r}$ is mapped back to the reference space through the consecutive upsampling and convolution operations using nearest-neighbor interpolation. Throughout the model, we use a filter size of $(3 \times 3)$ with the stride of one for convolution layers and $(2 \times 2)$ max pooling and upsampling operations with the same stride. We employ the hyperbolic-tangent (tanh) function $\varphi(z) = (e^z - e^{-z})/(e^z + e^{-z})$ as the non-linear activation function, since it led to the best performance in our study.

\subsection{CNN-based hierarchical autoencoders}
\begin{figure}[h]
    \centering
    \includegraphics[width=\textwidth]{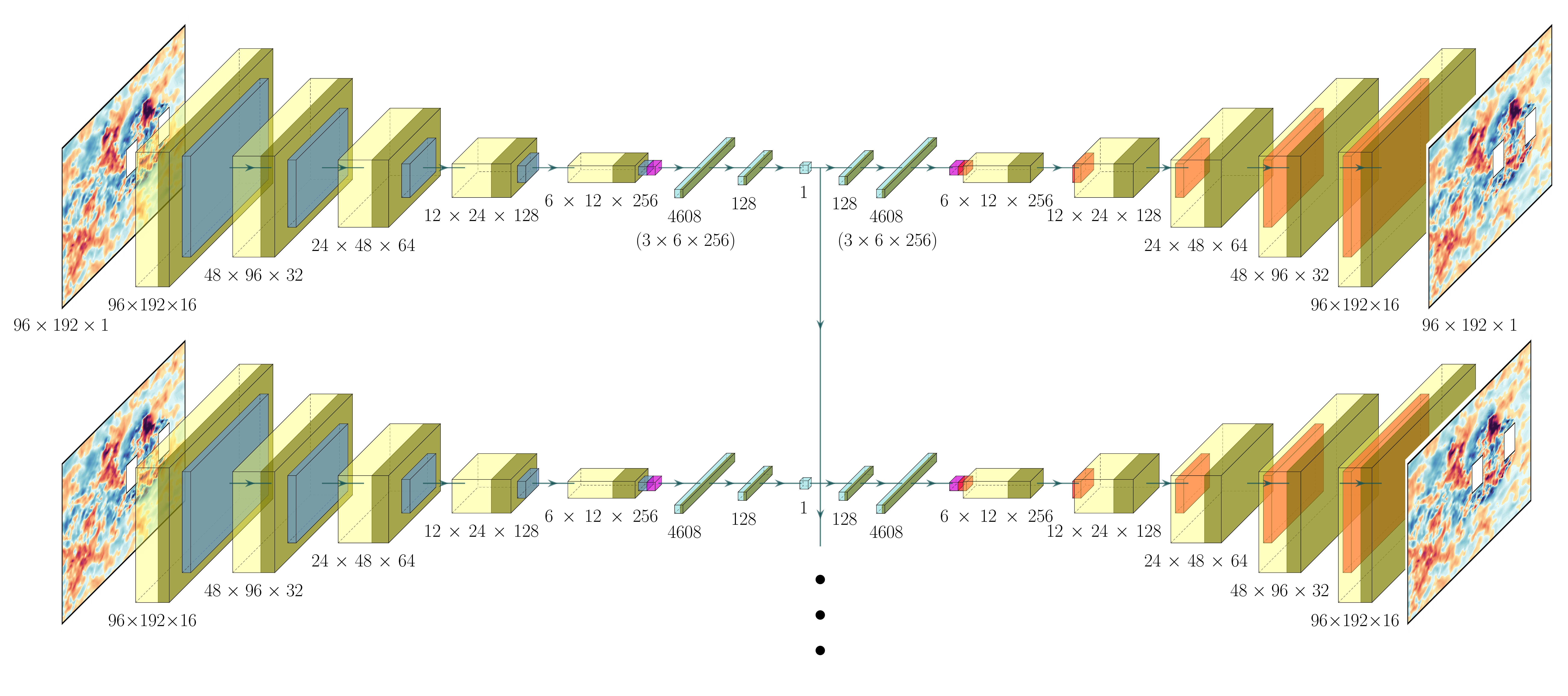}
    \caption{Schematic view of the CNN-HAE architectures with the same color coding as in \cref{fig:sch_AE}.}
    \label{fig:sch_HAE}
\end{figure}

Based on the idea proposed by \cite{SAEGUSA_2004}, \cite{Fukami_2020} proposed a CNN-based hierarchical autoencoder for modal decomposition of fluid flows in order to extract the modes ranked in terms of their contribution to the reconstruction while achieving more efficient data compression. To this end, the first subnetwork $\mathcal{F}_1$ is trained to map the high-dimensional data to a latent vector with size $d = 1$. The latent vector can be obtained using the encoder part of the first subnetwork as $\boldsymbol{r}_1 = \mathcal{E}_1(\boldsymbol{x})$. The second subnetwork $\mathcal{F}_2$ is then trained to reconstruct the input data at the output from a two-dimensional latent vector comprising the first latent vector $\boldsymbol{r}_1$, which has already been obtained, and the second latent vector $\boldsymbol{r}_2$, being updated through the training of $\mathcal{F}_2$, as $[\boldsymbol{r}_1~\boldsymbol{r}_2]$. The third network is trained in a similar way where the output is reconstructed from a latent vector as $[\boldsymbol{r}_1~\boldsymbol{r}_2~\boldsymbol{r}_3]$. A schematic of the CNN-HAE architecture is illustrated in \cref{fig:sch_HAE}. In the present study, we employ the same hyperparameters as those of the CNN-AE model for the CNN-HAE.

\subsection{CNN-based $\beta$-variational autoencoders}

In this section, we propose a modified version of the so-called variational autoencoders (VAEs) \citep{kingma2014,rezende_2014}, a probabilistic generative neural architecture, for modal decomposition of turbulent flows. VAEs are powerful generative models emerging from the combination of statistics and information theory with the flexibility of DNNs to efficiently generate new high-dimensional data. The goal is to map the data to a latent distribution, from which new meaningful samples can be generated. VAEs have gained increasing attention in the scientific community \citep{Iten_2020,Maulik_2020}, both due to their strong probabilistic foundation and their precious application in the field of representation learning \citep{Bengio_2013}.

\subsubsection{Marginal likelihood}
Let us consider a data sample $\boldsymbol{x}$ in some high-dimensional space $\mathcal{X}$ with the distribution $p(\boldsymbol{x})$. Let us assume that we have a vector of latent variables $\boldsymbol{r}$ in a low-dimensional space $\mathcal{R}$ with the probability density function (PDF) $p(\boldsymbol{r})$, from which we can easily sample new datapoints. Considering a family of deterministic functions $f(\boldsymbol{r};\boldsymbol{\theta})$, parameterized by a vector $\boldsymbol{\theta}$ in some space $\Theta$ in such a way that $f: \mathcal{R} \times \Theta \mapsto \mathcal{X}$, $f(\boldsymbol{r};\boldsymbol{\theta})$ is a random variable in the space of $\mathcal{X}$ if $\boldsymbol{r}$ is random and $\boldsymbol{\theta}$ is fixed. This implies that by optimizing $\boldsymbol{\theta}$ to maximize the probability of each $\boldsymbol{x}$, we can sample $\boldsymbol{r}$ from $p(\boldsymbol{r})$ and expect that $f(\boldsymbol{r};\boldsymbol{\theta})$ resembles the samples $\boldsymbol{x}$ in our dataset. Therefore, we aim to maximize:
\begin{equation}
    p_{\boldsymbol{\theta}}(\boldsymbol{x}) = \int p_{\boldsymbol{\theta}}(\boldsymbol{x}|\boldsymbol{r})p(\boldsymbol{r})\mathrm{d}\boldsymbol{r},
    \label{eq:max_likelihood}
\end{equation}
where $f(\boldsymbol{r};\boldsymbol{\theta})$ has been replaced by a distribution $p_{\boldsymbol{\theta}}(\boldsymbol{x}|\boldsymbol{r})$ to show the dependence of $\boldsymbol{x}$ on $\boldsymbol{r}$ according to the law of total probability, and $p_{\boldsymbol{\theta}}(\boldsymbol{x})$ is the so-called marginal likelihood, which is the approximation of $p(\boldsymbol{x})$ with parameters $\boldsymbol{\theta}$. This is the so-called maximum likelihood estimation (MLE), and the distribution $p(\boldsymbol{r})$ is often called the prior distribution. However, \cref{eq:max_likelihood} is typically intractable due to the integral and it is usually computationally infeasible \citep{kingma2014}. 

\subsubsection{Evidence lower bound (ELBO)}
VAEs define another probability distribution $q_{\boldsymbol{\phi}}(\boldsymbol{r}|\boldsymbol{x})$, a so-called probabilistic encoder (or recognition model). In a similar vein $p_{\boldsymbol{\theta}}(\boldsymbol{x}|\boldsymbol{r})$ is referred to as a probabilistic decoder (or generative model). The marginal likelihood is obtained as a sum over the marginal likelihoods of individual datapoints:
\begin{equation}
    \mathrm{log}\, p_{\boldsymbol{\theta}} ( \boldsymbol{x}^{(1)}, \dots, \boldsymbol{x}^{(m)}) = \sum_{i = 1}^m \mathrm{log}\, p_{\boldsymbol{\theta}}(\boldsymbol{x}^{(i)}), 
\end{equation}
and for each $\boldsymbol{x}$ it can be defined as:
\begin{equation}
    \mathrm{log}\, p_{\boldsymbol{\theta}}(\boldsymbol{x}) = D_{\mathrm{KL}}(q_{\boldsymbol{\phi}}(\boldsymbol{r}|\boldsymbol{x})||p_{\boldsymbol{\theta}}(\boldsymbol{r}|\boldsymbol{x})) + \boldsymbol{\mathcal{C}}(\boldsymbol{\theta}, \boldsymbol{\phi}; \boldsymbol{x}),
\end{equation}
where the first right-hand-side (RHS) term is the Kullback--Leibler (KL) divergence $D_{\mathrm{KL}}$ between $q_{\boldsymbol{\phi}}(\boldsymbol{r}|\boldsymbol{x})$ and $p_{\boldsymbol{\theta}}(\boldsymbol{r}|\boldsymbol{x})$. The KL-divergence is non-negative, which indicates that the second RHS term $\mathcal{C}(\boldsymbol{\theta}, \boldsymbol{\phi}; \boldsymbol{x})$ is a lower bound on the marginal likelihood, and can be written as:
\begin{equation}
    \mathrm{log}\, p_{\boldsymbol{\theta}}(\boldsymbol{x}) \geq \boldsymbol{\mathcal{C}}(\boldsymbol{\theta}, \boldsymbol{\phi}; \boldsymbol{x}) = - D_{\mathrm{KL}}(q_{\boldsymbol{\phi}}(\boldsymbol{r}|\boldsymbol{x})||p_{\boldsymbol{\theta}}(\boldsymbol{r})) + \mathbb{E}_{q_{\boldsymbol{\phi}}(\boldsymbol{r}|\boldsymbol{x})} \left[ \mathrm{log}\, p_{\boldsymbol{\theta}}(\boldsymbol{x}|\boldsymbol{r}) \right].
    \label{eq:ELBO}
\end{equation}
This term is usually called evidence lower bound (ELBO). Since the ELBO is more tractable than the MLE, it is used as the cost function for the training of NNs to optimize the unknown parameters $\boldsymbol{\theta}$ and $\boldsymbol{\phi}$. 

\subsubsection{Reparameterization trick}
Since the operation that samples a latent vector from $q_{\boldsymbol{\phi}(\boldsymbol{r}|\boldsymbol{x})}$ is not differentiable, we need to perform a change of variable, the so-called reparameterization trick \citep{kingma2014}, to differentiate ELBO with respect to both $\boldsymbol{\theta}$ and $\boldsymbol{\phi}$. We assume $q_{\boldsymbol{\phi}}(\boldsymbol{r}|\boldsymbol{x})$ to be a Gaussian distribution,
\begin{equation}
    \mathrm{log}\, q_{\boldsymbol{\phi}(\boldsymbol{r}|\boldsymbol{x})} = \mathrm{log}\, \boldsymbol{\mathcal{N}}(\boldsymbol{r}; \boldsymbol{\mu}, \boldsymbol{\sigma}^2\mathbf{I}),
\end{equation}
where the mean $\boldsymbol{\mu}$ and the standard deviation $\boldsymbol{\sigma}$ are outputs of the encoder, and $\mathbf{I}$ is the identity matrix. We sample from $q_{\boldsymbol{\phi}}(\boldsymbol{r}|\boldsymbol{x})$ using $\boldsymbol{r} = \boldsymbol{\mu} + \boldsymbol{\sigma} \odot \boldsymbol{\varepsilon}$ where $\boldsymbol{\varepsilon} \sim  \boldsymbol{\mathcal{N}}(\mathbf{0}, \mathbf{I})$ is an auxiliary normally-distributed random number, and $\odot$ indicates an element-wise product. Moreover, the term $\mathbb{E}_{q_{\boldsymbol{\phi}}(\boldsymbol{r}|\boldsymbol{x})} \left[ \mathrm{log}\, p_{\boldsymbol{\theta}}(\boldsymbol{x}|\boldsymbol{r}) \right]$, which is the so-called log-likelihood, encourages accurate reconstruction of the data and can be estimated as a negative reconstruction error in an autoencoder setting \citep{kingma2014}. This leads to the VAE cost function $\mathcal{C}$, and we can take the negative of it as a loss function $\mathcal{L}$ for training the NNs:
\begin{equation}
    \mathcal{L}(\boldsymbol{\theta}, \boldsymbol{\phi}; \boldsymbol{x}) = \boldsymbol{\mathcal{L}}_{\mathrm{rec}} - \dfrac{1}{2} \sum_{i = 1}^{d}(1 + \mathrm{log}(\sigma_i^2) - \mu_i^2 - \sigma_i^2).
    \label{eq:loss_VAE}
\end{equation}

\subsubsection{Disentangled representation}
In the field of representation learning \citep{Bengio_2013}, it is of interest to find a latent representation of the high-dimensional data as an uncorrelated representation with a minimal number of parameters (factors), the so-called disentangled representation, which can be useful for a large variety of tasks and domains.~\cite{Higgins_2017} proposed to augment the original VAE loss function with a single hyperparameter $\beta \geq 0$ that controls the extent of the learning constraints. The goal is to encourage learning of statistically-independent latent variables $\boldsymbol{r}_i$ and penalize the size of the latent vector $\boldsymbol{r}$. This can be obtained by minimizing the distance $D_{\mathrm{KL}} \left [ p(\boldsymbol{r})||\prod_i p(\boldsymbol{r}_i) \right ]$ between $p(\boldsymbol{r})$ and the product of its marginals. In practice, this is performed by upweighting the KL term in the ELBO, see \cref{eq:ELBO}, with a penalization factor $\beta$ leading to the following loss function:
\begin{equation}
    \mathcal{L}(\boldsymbol{\theta}, \boldsymbol{\phi}; \boldsymbol{x}) = \boldsymbol{\mathcal{L}}_{\mathrm{rec}} - \dfrac{\beta}{2} \sum_{i = 1}^{d}(1 + \mathrm{log}(\sigma_i^2) - \mu_i^2 - \sigma_i^2)
    \label{eq:loss_BVAE}
\end{equation}
for $\beta$-VAEs. A detailed discussion on disentangling in $\beta$-VAEs can be found in the works by \cite{Higgins_2017,burgess_2018,Achille_2018,locatello2019}.

\begin{figure}[h]
    \centering
    \includegraphics[width=\textwidth]{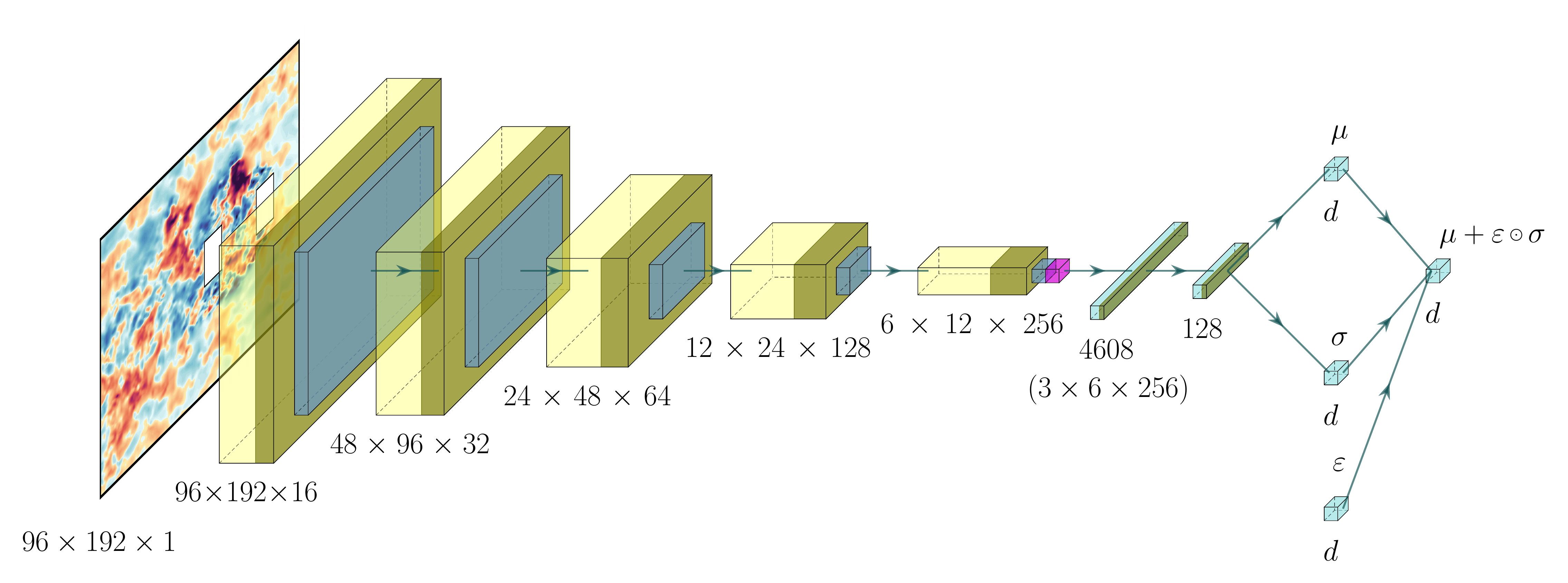}
    \caption{Schematic view of the encoder part $\mathcal{E}$ of the CNN-$\beta$VAE with the same color coding as in \cref{fig:sch_AE}.}
    \label{fig:sch_VAE}
\end{figure}

In the present work, we employ a CNN-based $\beta$-VAE architecture for modal decomposition of turbulent flows. Our goal is to minimize the correlation between the latent variables, motivating the network to extract a set of orthogonal modes, and also penalize the size of the latent vector $d$. This leads to an efficient representation of the high-dimensional data useful for flow analysis, reduced-order modeling, and flow control.~\Cref{fig:sch_VAE} illustrates the schematic of the encoder part of the CNN-$\beta$VAE model. The decoder part is the same as the decoder part of the CNN-AE model depicted in \cref{fig:sch_AE}.

\section{Results and discussion}
\label{sec:results-and-discussion}

The key insight of the present study is to encourage independence of the latent variables $\boldsymbol{r}_1, \dots, \boldsymbol{r}_d$ to extract near-orthogonal modes from turbulent flows using CNN-$\beta$VAEs. This is to motivate disentangled representations in the language of representation learning. We impose a limit on the capacity of the latent information and motivate learning statistically independent latent variables using the penalization factor $\beta$. The objective is to motivate orthogonality in the latent space to obtain modes that are useful for flow analysis, reduced-order modeling, and flow control.

The KL term in the $\beta$-VAE loss function, see \cref{eq:loss_BVAE}, acts as a regularizer. This regularization may create a trade-off between reconstruction fidelity and the quality of learning independent representations. In this section, we investigate the effect of the penalization factor $\beta$ and the size of the latent vector $d$ on the performance of the CNN-$\beta$VAE models, both in terms of reconstruction accuracy and independence of the modes. Furthermore, we compare the performance of the CNN-$\beta$VAE in the modal decomposition of the turbulent flow through a simplified urban environment with that of the CNN-HAE, CNN-AE, and POD. To this end, we define two evaluation metrics to measure the quality of the reconstructions and the orthogonality (disentanglement) of the latent variables. For reconstruction quality, we evaluate the turbulent-kinetic-energy percentage $E_k$ that is captured by the model reconstructions as:
\begin{equation}
    E_k = \left ( 1 - \left\langle \dfrac{\sum \limits_{i = 1}^{n} \left(u - \tilde{u}\right)^2}{\sum \limits_{i = 1}^{n}u^2} \right\rangle \right ) \times 100,
\end{equation}
where $\langle \cdot \rangle$ indicates ensemble averaging in time, $u$ and $\tilde{u}$ denote the reference value of the fluctuating component of streamwise velocity and its reconstruction, respectively, and $n$ is the number of grid points. To measure the independency of the latent variables, we compute the determinant of the correlation matrix multiplied by 100 and refer to it as $\mathrm{det}_{\mathbf{R}}$, where $\mathbf{R} = (R_{ij})_{d \times d}$ is the correlation matrix defined by:
\begin{equation}
    R_{ii} = 1~\mathrm{and}~R_{ij} = \dfrac{C_{ij}}{\sqrt{C_{ii} C_{jj}}},
\end{equation}
for all $1 \leq i \neq j \leq d$, and $C_{ij}$ denotes the components $i, j$ of the covariance matrix $\mathbf{C}$. Note that $\mathrm{det}_{\mathbf{R}}$ is 100 when all the variables are completely uncorrelated ($R_{ij} = 0$) and zero when they are completely correlated ($R_{ij} = 1$). We report the value of $\mathrm{det}_{\mathbf{R}}$ as a metric for independency of the latent variables.

\subsection{Learning orthogonal and parsimonious modes}
\label{sec:bvae_results}
The introduction of non-linearity in the process of modal decomposition using AEs through the use of non-linear activation functions leads to excellent performance in terms of reconstruction accuracy in comparison to the linear-theory-based decomposition methods such as POD. However, the AE-based modes lack the useful properties of the linear-theory-based modes such as orthogonality and ranking, which may lead to interpretability. We constrain the shape of the latent space to motivate orthogonality (disentanglement) of the features extracted by the CNN-$\beta$VAEs and obtain parsimonious (minimal) modes. The penalization factor $\beta$ regulates the balance between the information preservation and the information capacity of the latent vector \citep{Higgins_2017}.

\begin{figure}[th]
    \centering
    \begin{overpic}[width=\textwidth]{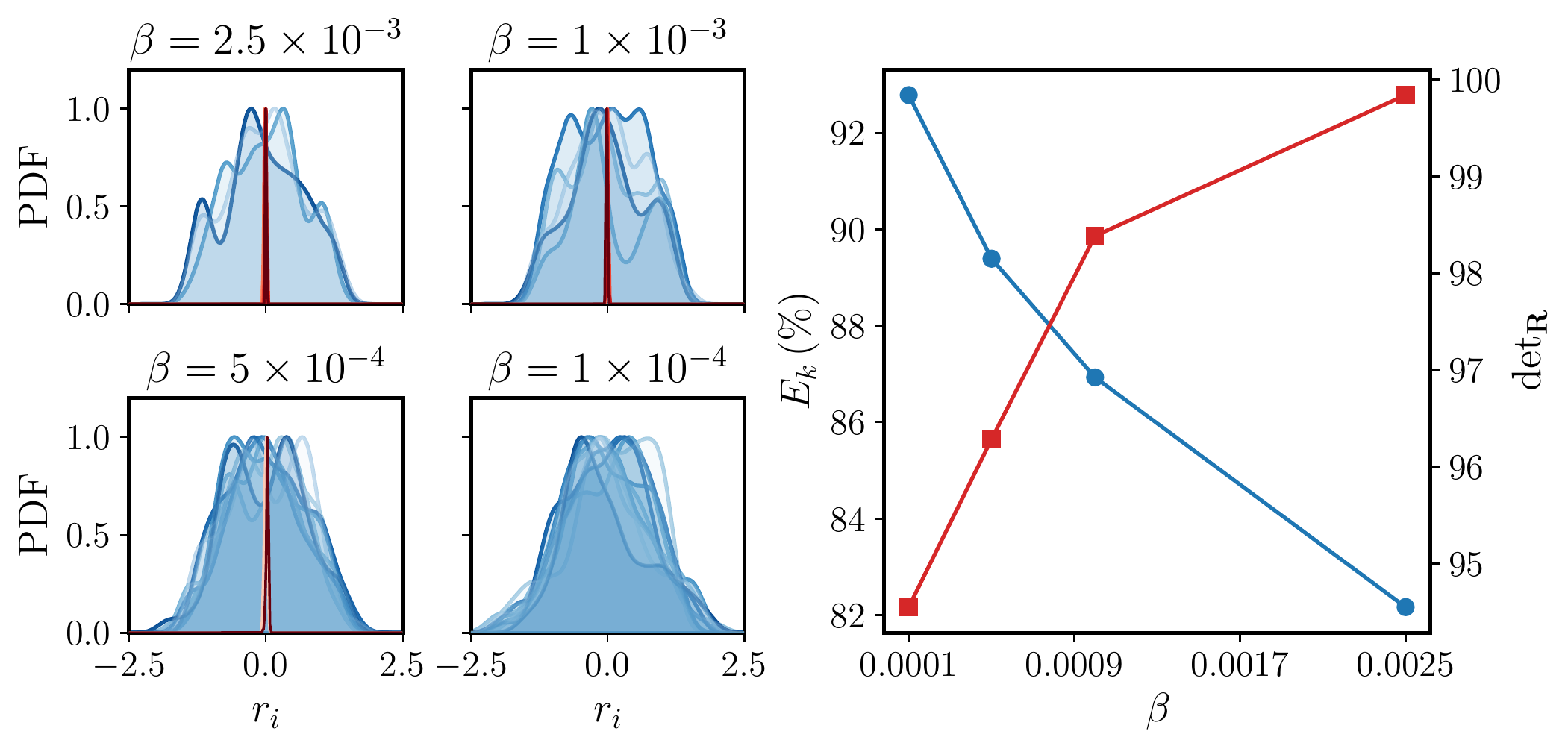}
        \put (0,45) {(a)}
        \put (50,45) {(b)}
    \end{overpic}
    \caption{Effects of the penalization factor $\beta$ on the performance of the CNN-$\beta$VAE models. (a) Normalized probability density function (PDF) of the latent variables obtained from models with different values of $\beta$ as indicated in each panel. The variables with considerably large values are colored by blue and the variables with very small deviations from zero are depicted by red. (b) Effect of $\beta$ on $E_k$ (blue) and $\mathrm{det}_\mathbf{R}$ (red).}
    \label{fig:pdfs}
\end{figure}

The results obtained from CNN-$\beta$VAEs using four different values of $\beta$ are reported in \cref{fig:pdfs}. For these tests we consider $d = 10$ and $\beta = 10^{-4}$, $5 \times 10^{-4}$, $10^{-3}$, and $2.5 \times 10^{-3}$. \Cref{fig:pdfs}(a) depicts the normalized probability density function (PDF) obtained from each variable. We note that using $\beta = 2.5 \times 10^{-3}$, only three latent variables have significant values and the others are comparatively very small. These variables have a very small deviation from zero, and thus can be regarded as noise. We also observe that the spatial modes corresponding to these variables are very similar. Therefore, by imposing a limit on the capacity of the latent information using $\beta = 2.5 \times 10^{-3}$ the network minimizes the latent-space dimension to three and leads to almost perfectly independent variables where $\mathrm{det}_{\mathbf{R}} = 99.84$. This test also leads to a very good reconstruction of the turbulent flow with $E_k = 82.17\%$ as it is shown in \cref{fig:pdfs}(b). By reducing the penalization factor $\beta$ to $10^{-3}$, the latent-space dimension increases to five as a consequence of a lighter constrain on the capacity of the latent information, which also leads to a reduction of $\mathrm{det}_{\mathbf{R}}$ to 98.38. However, the reconstruction performance is improved leading to $E_k = 86.93\%$. By further reduction of $\beta$, the dimension of the latent space is increased where all the 10 latent variables have considerable values using $\beta = 10^{-4}$. Moreover, it can be observed in \cref{fig:pdfs}(b) that by reducing $\beta$, the independence of the variables diminishes while the reconstruction accuracy is improved. We obtain $\mathrm{det}_{\mathbf{R}} = 94.55$ and $E_k = 92.78\%$ using $\beta = 10^{-4}$.

Next, we investigate the effect of the size of the latent vector $d$ on the performance of the CNN-$\beta$VAEs. For these tests, we consider $\beta = 10^{-3}$ and $d = 5$, 10, 20. We observe that in all three tests, only five latent variables have considerable values and the rest are just noise. Our results show that although we increase $d$, the penalization using $\beta = 10^{-3}$ encourages the network to learn a minimal number of near-orthogonal modes (only five) for accurate reconstruction of the turbulent flow. Since the network maps the information into only five variables, increasing $d$ to more than five does not improve the performance of the model. We obtain $\mathrm{det}_{\mathbf{R}}$ equal to 99.20, 98.38, and 99.22 and $E_k$ equal to 87.36\%, 86.93\%, and 87.15\% for the CNN-$\beta$VAEs with $d$ equal to 5, 10, and 20, respectively.

\begin{figure}[h]
    \centering
    \vspace{10pt}
    \begin{overpic}[width=\textwidth]{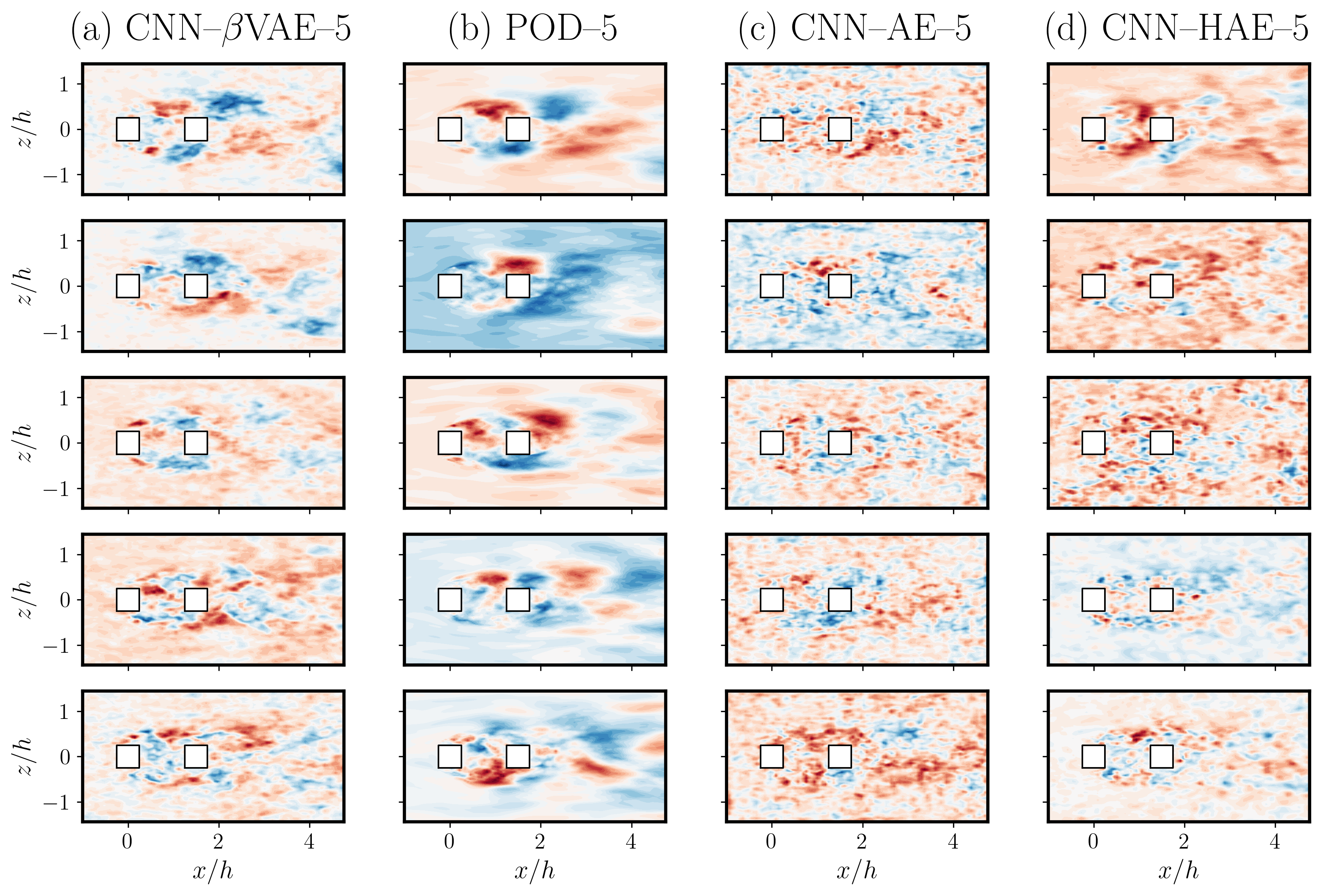}
        \put (-5,53.5) {\rotatebox{90}{mode 1}}
        \put (-5,41.5) {\rotatebox{90}{mode 2}}
        \put (-5,29.5) {\rotatebox{90}{mode 3}}
        \put (-5,17.5) {\rotatebox{90}{mode 4}}
        \put (-5,5.5) {\rotatebox{90}{mode 5}}
    \end{overpic}

    \caption{The ranked spatial modes obtained from the CNN-$\beta$VAE (a), POD (b), CNN-AE (c), and the CNN-HAE (d). The size of the latent vector $d$ is equal to 5.}
    \label{fig:ordered_modes}
\end{figure}
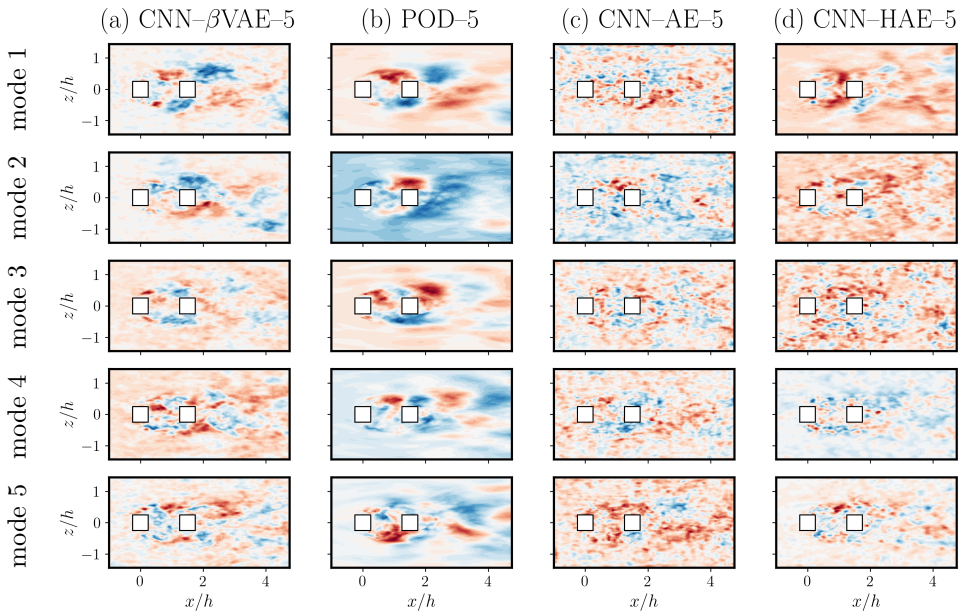

\subsection{Ranking the CNN-$\beta$VAE modes}

As we discussed in $\S$\ref{sec:POD}, POD modes are sorted in terms of their energy content. This property is extremely useful for understanding and analyzing the dominant patterns in complex flows. \cite{Fukami_2020} implemented hierarchical autoencoders to extract AE-based modes in the order of their contribution in the reconstruction, which requires training multiple NNs and might be cumbersome especially for the extraction of higher-order modes. Here, we propose a strategy for ordering the CNN-$\beta$VAE modes. Later, we show that CNN-$\beta$VAEs are able to extract near-orthogonal and parsimonious modes from turbulent flows. These properties allow us to rank these modes after the training process and based on their contribution to the reconstruction. In particular, we rank CNN-$\beta$VAE modes based on the maximum $E_k$ that can be obtained from $q$ modes, where $q$ represents the rank. To this end, after the training process, we first use the encoder to map high-dimensional data to the latent vector $\mathcal{E}: \boldsymbol{x} \mapsto \boldsymbol{r}$. We zero out all the latent variables except the $i^{th}$ variable, which leads to a latent vector $\hat{\boldsymbol{r}}_i$. Then, we employ the decoder part of the CNN-$\beta$VAE to send this latent vector to the original space $\mathcal{D}: \hat{\boldsymbol{r}}_i \mapsto \tilde{\boldsymbol{x}}_i$. This procedure is performed for all the time steps. The kinetic-energy percentage that is captured by only considering the $i^{th}$ mode is evaluated as $E_{k}^i$. The first mode is selected as the mode leading to the maximum value of $E_{k}^i$. For the second mode, we perform the same procedure while we preserve the first mode and look for the mode, which in combination with the first mode, leads to the maximum value of $E_k^i$. In a similar way, the third mode is selected as the mode which gives the maximum $E_k^i$ in combination with the first and second modes. We continue this procedure to rank all the modes.~\Cref{fig:ordered_modes} illustrates the ranked modes obtained from a CNN-$\beta$VAE model with $d = 5$ and $\beta = 10^{-3}$, together with the modes obtained from the POD, CNN-AE, and CNN-HAE methods with $d = 5$. The POD and the CNN-HAE modes are already ranked and we also perform the ordering procedure for the CNN-AE modes. A clear resemblance can be observed between the first two modes of the CNN-$\beta$VAE, see \cref{fig:ordered_modes}(a), and those from POD, as shown in \cref{fig:ordered_modes}(b), indicating the ability of CNN-$\beta$VAEs in the extraction of interpretable modes from turbulent flows. These modes correspond to the large-scale vortex shedding from around the obstacles into the wake region. Moreover, these results show that using the ranking procedure it is possible to sort the CNN-$\beta$VAE modes based on their importance for reconstruction. It also can be seen in \cref{fig:ordered_modes}(c) that it is extremely difficult to relate CNN-AE modes to physical processes, a fact that is referred to as the lack of interpretability. We observe that although the CNN-HAE model is able to extract large-scale features first, the obtained modes may not be physically interpretable, as shown in \cref{fig:ordered_modes}(d).

\begin{figure}[h]
    \centering
    \includegraphics[width=\textwidth]{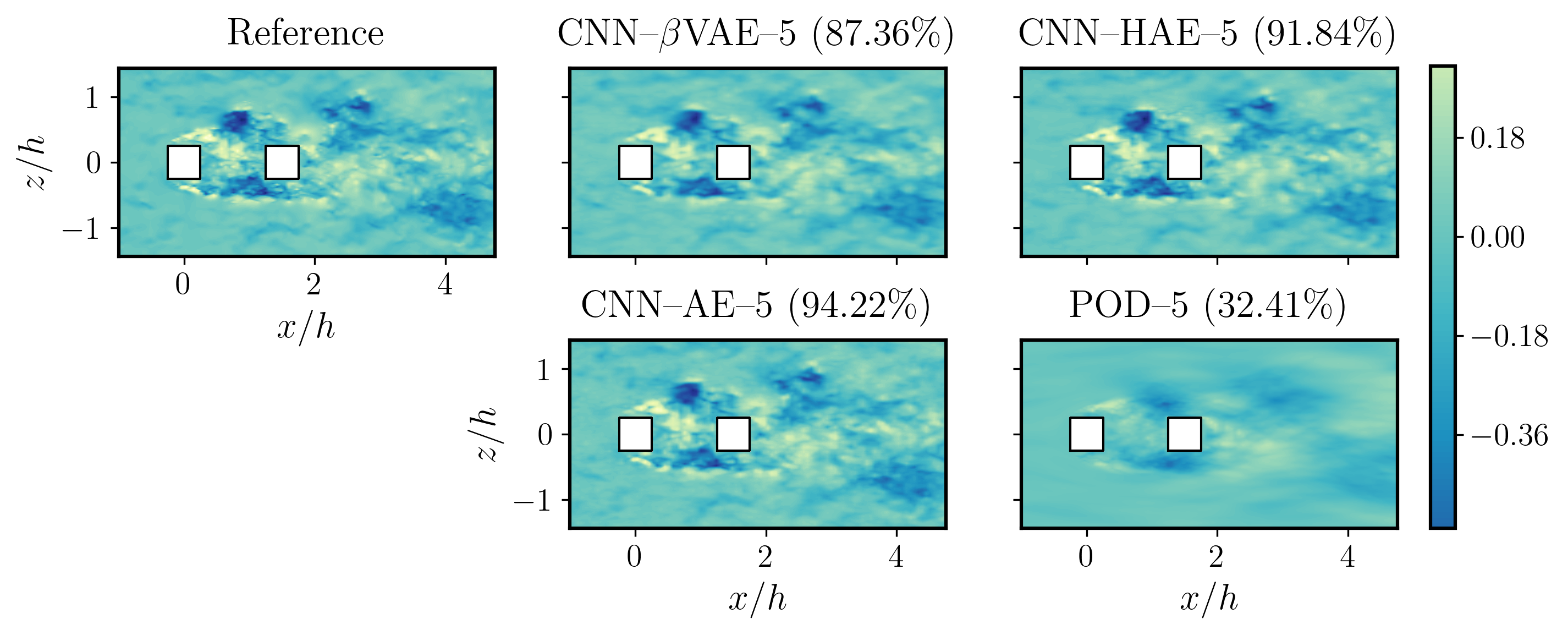}
    \caption{Reconstruction of the fluctuating component of the streamwise velocity obtained from different methods, as indicated in each panel, in comparison with the reference data. The value in brackets on each panel indicates the obtained $E_k$. 
    }
    \label{fig:reconstructions}
\end{figure}

\begin{figure}[h]
    \centering
    \includegraphics[width=\textwidth]{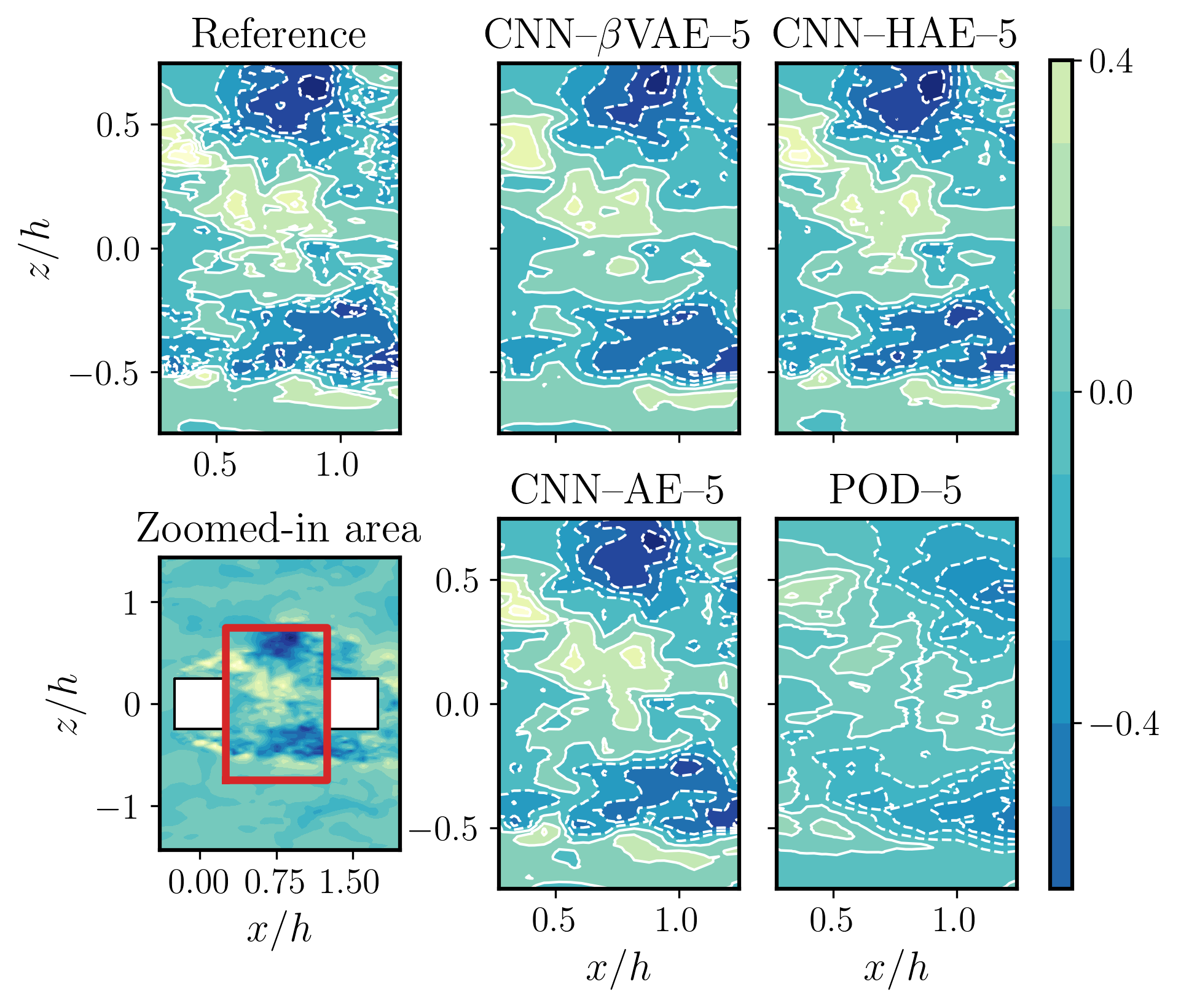}
    \caption{Reconstruction of the fluctuating component of the streamwise velocity obtained from different methods, as indicated on each panel, in comparison with the reference data for the zoomed-in area between the two obstacles marked by the red rectangle. 
    }
    \label{fig:reconstructions_zoom}
\end{figure}

\subsection{Comparison of CNN-AE-based models}

In previous sections, we proposed CNN-$\beta$VAEs for modal decomposition of turbulent flows and showed their ability in the extraction of near-orthogonal and parsimonious modes. Here, we compare the performance of CNN-$\beta$VAEs with that of CNN-HAEs, CNN-AEs and POD in terms of reconstruction accuracy and orthogonality of the latent variables. We select $d = 5$ as the size of the latent vector for all models. For the CNN-$\beta$VAE we use $\beta = 10^{-3}$. \Cref{fig:reconstructions} shows the reconstruction of the first time step in the dataset obtained from different methods in comparison with the reference data. It can be observed that a more accurate reconstruction can be obtained using NN-based methods in comparison with 5 POD modes due to the introduction of non-linearity in the algorithm. The CNN-AE model leads to the best reconstructions with $E_k = 94.22 \%$ while 32.41\% of the kinetic energy is captured by the 5 POD modes. Both CNN-HAE and CNN-$\beta$VAE models also lead to excellent reconstructions with $E_k$ of 91.84\% and 87.36\%, respectively, which are slightly lower than that of the CNN-AE. For the CNN-$\beta$VAE, it is due to the fact that the regularization with the KL term in the $\beta$-VAE loss function, \cref{eq:loss_BVAE}, induces a trade-off between the reconstruction quality and learning independent representations. However, our results show that excellent reconstructions can be obtained using all the three CNN-AE-based models leading to $E_k$ of about 90\% using only 5 modes. 

The reconstruction results are also reported in a detailed area between the two obstacles in \cref{fig:reconstructions_zoom} to provide a clear insight into the fidelity of the reconstructions. It can be seen that although some small-scale features are lost, all three CNN-AE models are able to preserve the dominant structures of the turbulent flow. However, POD can not reconstruct the turbulent flow properly from 5 modes.

\begin{figure}[h]
    \centering
    \includegraphics[width=\textwidth]{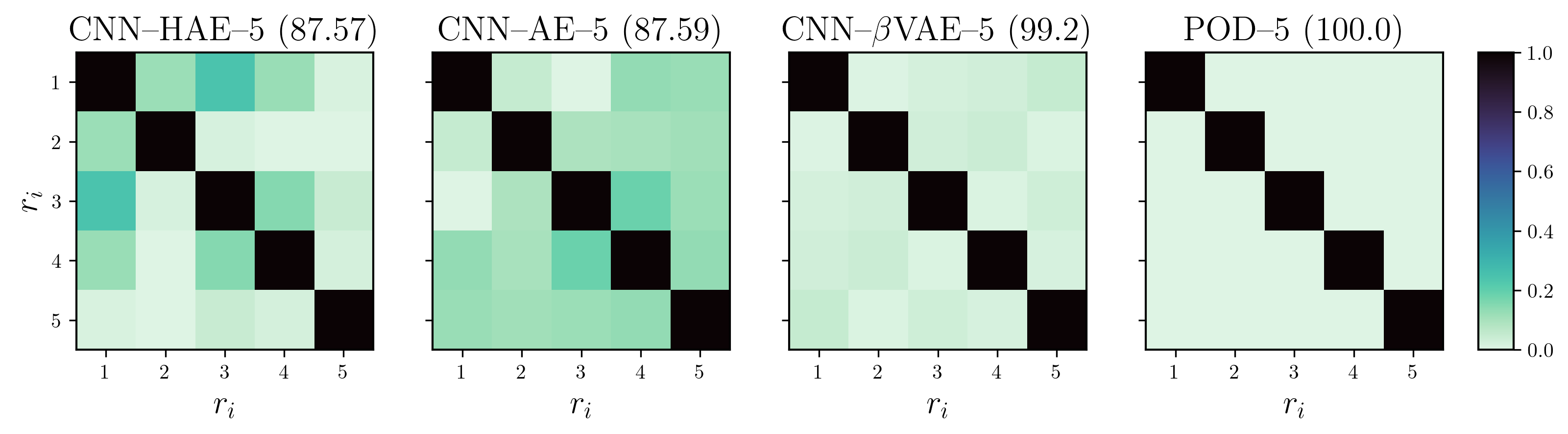}
    \caption{Correlation matrix $\mathbf{R}$ for the latent variables obtained from different models as indicated on the panels. The value in brackets indicates the corresponding $\mathrm{det}_\mathbf{R}$ for each case.}
    \label{fig:corrcoefs}
\end{figure}

Next, we compare the independence of the latent variables obtained from different methods. As we discussed in $\S$\ref{sec:POD} and $\S$\ref{sec:bvae_results}, orthogonality of the modes is a useful property for flow analysis, reduced-order modeling and flow control. Moreover, motivating the orthogonality of the modes may lead to interpretability. Results are depicted in \cref{fig:corrcoefs} as the absolute value of the correlation matrix $\mathbf{R}$ corresponding to the latent variables from the CNN-HAE, CNN-AE, CNN-$\beta$VAE with $d = 5$, and also POD with 5 modes. It can be seen that although the CNN-HAE model extract modes in the order of their contribution in the reconstruction, the latent variables are correlated leading to the lowest value for $\mathrm{det}_{\mathbf{R}}$ among all methods. As mentioned above, it is possible to motivate the disentanglement or independence of the latent variables using CNN-$\beta$VAEs and obtain near-orthogonal modes. It can be observed that the correlation between the latent variables is reduced for the CNN-$\beta$VAE in comparison to that of the CNN-AE, where $\mathrm{det}_{\mathbf{R}}$ is equal to 99.20 for the CNN-$\beta$VAE method and 87.59 for the CNN-AE technique.

\section{Summary and conclusions}
\label{sec:conclusions}

In this study, we proposed a probabilistic deep-neural-network architecture based on $\beta$-VAEs and CNNs for non-linear mode decomposition of turbulent flows. The objective is to learn a compact, near-orthogonal and parsimonious latent representation of high-dimensional data by introducing non-linearity in the process of low-dimensionalization and also minimizing the correlation between the latent variables and penalizing the size of the latent vector, which may lead to a set of interpretable modes useful for flow analysis, reduced-order modeling, and flow control. Since the correlations among the learned latent variables are minimized, we proposed an algorithm to rank the VAE-based modes based on their contribution to the reconstruction. We applied the proposed CNN-$\beta$VAE architecture for modal decomposition of the turbulent flow through a simplified urban environment. Furthermore, we compared the performance of the CNN-$\beta$VEs in terms of the quality of the reconstructions and orthogonality of the extracted modes with that of the CNN-AEs and CNN-HAEs. The flow-field data were obtained from a well-resolved large-eddy simulation (LESs) database~\cite{Torres2021,TorresMS}. Our results from modal decomposition using POD indicates that 247 modes are required to obtain 99\% of the energy from the reconstruction. This indicates that it is challenging to represents turbulent flows as a linear superposition of a few POD modes. Our proposed CNN-$\beta$VAE model with a latent vector size $d = 5$ and a penalization factor of $\beta = 10^{-3}$ leads to $E_k$ of 87.36\% against 32.41\% obtained from POD, which shows the excellent performance of the CNN-$\beta$VAE in the reconstruction of the turbulent flow from only five modes. This model also leads to near-orthogonal modes, where $\mathrm{det}_{\mathbf{R}}$ is equal to $99.20$. We showed that by constraining the shape of the latent space and motivating orthogonality of the modes, we can extract meaningful non-linear features where the first mode of this CNN-$\beta$VAE model represents the large-scale vortex shedding from around the obstacles into the wake. Moreover, we investigated the effects of the size of the latent vector $d$ and the penalization factor $\beta$ on the performance of the model. Our results indicate the ability of the CNN-$\beta$VAEs in learning a minimal (parsimonious) set of near-orthogonal modes which are required for reconstruction, where the penalization encourages the model to minimize the number of learned latent variables and it does not change by increasing $d$. Moreover, we observed that there is a trade-off between the reconstruction accuracy and the quality of learning independent representations. Our results showed that a lighter constrain on the capacity of the latent information leads to a slightly better reconstruction while allowing more correlations among the latent variables. Our comparison between the CNN-$\beta$VAE, CNN-HAE, and CNN-AE models indicates that although motivating orthogonality of the modes decreases the reconstruction accuracy, very good reconstructions can be obtained from five modes using the CNN-$\beta$VAE leading to $E_k$ of 87.36\% against 93.93\% and 91.84\% of the CNN-AE and the CNN-HAE, respectively. The CNN-$\beta$VAE model leads to a set of near-orthogonal modes with the highest $\mathrm{det}_{\mathbf{R}}$ among the AE-based models. The proposed CNN-$\beta$VAE architecture can be extended in future works for the development of reduced-order surrogate models or advanced flow-control methods, among others.

\section*{Acknowledgments}
We acknowledge \'Alvaro Mart\'inez for his contributions to this work. RV acknowledges the G\"oran Gustafsson foundation for the financial support of this research. SH has been supported by project RTI2018-102256-B-I00 of Mineco/FEDER. SLC acknowledges the support of the Spanish Ministry of Science and Innovation under the grant PID2020-114173RB-100.

\bibliographystyle{elsarticle-harv}
\biboptions{authoryear}
\bibliography{References}

\end{document}